\documentclass[12pts,twocolumn,epsf,aps,prb]{revtex4-1}
\usepackage{float}
\usepackage{color,graphicx}
\usepackage[colorlinks=true,linkcolor=blue,filecolor=blue,urlcolor=blue]{hyperref}

\begin{document}

\title{Giant magnetocaloric effect in exchange-frustrated GdCrTiO$_5$ antiferromagnet}

\author{M. Das, S. Roy, N. Khan, and P. Mandal}

\affiliation{Saha Institute of Nuclear Physics, HBNI, 1/AF Bidhannagar, Calcutta 700064, India.}
\date{\today}

\begin{abstract}

We report the effect of exchange frustration on the magnetocaloric properties of GdCrTiO$_5$ compound. Due to the highly exchange-frustrated nature of magnetic interaction, in GdCrTiO$_5$, the long-range antiferromagnetic ordering occurs at much lower temperature $T_N$=0.9 K and the magnetic cooling power enhances dramatically relative to that observed in several geometrically frustrated systems. Below 5 K, isothermal magnetic entropy change (-$\Delta S_{\rm m}$) is found to be 36 J kg$^{-1}$ K$^{-1}$, for a field change ($\Delta H$) of 7 T. Further, -$\Delta S_{\rm m}$ does not decrease from its maximum value with decreasing in $T$ down to very low temperatures and is reversible in nature. The adiabatic temperature change, $\Delta T_{\rm ad}$, is 15 K for $\Delta H$=7 T. These magnetocaloric parameters are significantly larger than that reported for several potential magnetic refrigerants, even for small and moderate field changes. The present study not only suggests  that GdCrTiO$_5$ could be considered as a potential magnetic refrigerant at cryogenic temperatures but also promotes further studies on the role of exchange frustration on  magnetocaloric effect. In contrast, only the role of geometrical frustration on magnetocaloric effect has been previously reported theoretically and experimentally investigated on very few systems.
\end{abstract}
\maketitle

\textbf{I. INTRODUCTION }

The potential technological applications of environment friendly magnetic refrigeration technique such as in space science, liquefaction and storage of hydrogen in the fuel industry and to achieve sub-kelvin temperatures for basic research seek for the most efficient low-temperature magnetocaloric material\cite{gsc,gut,shen,yu,liu,te,fol,pec,vk,ze,von,caro,li,ros,ba,rm,br,wa,mil}. Paramagnetic salts are the standard refrigerant materials to achieve the sub-kelvin temperatures using adiabatic demagnetization techniques. The high density and very large magnetic moments in a magnetic refrigerant enhance the magnetic cooling power\cite{fol,pec,vk,ze,von,caro,pam}. Several rare-earth element based transition metal oxides and intermetallic compounds carrying high magnetic moments have become attractive candidates for the low-temperature magnetic refrigeration.\cite{prosen,ke,ji,ycao,hua,sh,midya,lo,bali,sp} In these materials, the rare-earth magnetic moments order at low temperature and a strong suppression of the magnetic entropy takes place in the vicinity of the order-disorder phase transition with the application of magnetic field. However, the value of the magnetic entropy, $\Delta S_{\rm m}$, decreases rapidly and becomes very small just few Kelvin below the transition temperature and thereby limits the lowest temperature achievable by the magnetic refrigeration technique. This is one of the major drawbacks for refrigeration using magnetically ordered materials. Recently, it has been shown that the presence of magnetic frustration significantly enhances the magnetic cooling power with finite residual magnetic entropy well below the Neel temperature.\cite{midya,bali,sp}  Frustration leads to infinite degeneracy of the magnetic ground state which implies the presence of a macroscopic number of local zero-energy modes in the system in zero-field. Above the saturation field $H_{\rm sat}$, a nondegenerate fully polarized spin state of the antiferromagnet is achieved. Adiabatic demagnetization of this state corresponds to the condensation of macroscopic number of local zero-energy modes and thereby produces a large change in magnetic entropy leading to magnetic cooling of the system. The role of frustration on the magneto-caloric effect has been investigated theoretically for classical Heisenberg antiferromagnets on different geometrically frustrated lattices such as kagome, garnet, and pyrochlore lattices\cite{me,ss}. It has been observed that the pyrochlore lattices being the most frustrated among the above three lattices and offers the fastest cooling rate under adiabatic demagnetization among geometrically frustrated magnets.\cite{me,ss} In the present experimental study, we investigate the magnetocaloric effect of highly frustrated GdCrTiO$_5$ compound where the competing magnetic exchange interactions from two sublattices is the origin of the magnetic frustration. The role of frustration, which is due to the bond dependent anisotropic exchange interactions or competing magnetic exchange interactions, on the magentocaloric effect should also be explored for a complete understanding of the frustration induced enhancement of the magnetic cooling power.\\

The rare-earth compounds of the type $R$Mn$_2$O$_5$, crystallizing in an orthorhombic structure (space group Pbam), have attracted a lot of attention due to their magnetoelectric coupling and magnetic-field induced ferroelectric behavior. Besides  magnetoelectric properties, $R$Mn$_2$O$_5$ shows the ability to be a good magnetic refrigerant at cryogenic temperature.\cite{bali, gr, is, wak, jm} However, the members of the family $R$CrTiO$_5$ isostructural to $R$Mn$_2$O$_5$  were paid very little attention. \cite{sa, mg, hw} In orthorhombic $R$CrTiO$_5$, the Cr$^{3+}$  ions are interspaced between the $R$$^{3+}$ and Ti$^{4+}$ ions  and the Cr$^{3+}$ spins are collinear along the crystallographic $c$ axis whereas the  moments of $R$$^{3+}$ lie on the $ab$ plane.\cite{tb, bas} The schematic crystal structure of GdCrTiO$_5$ is shown in Fig. \textbf{1} and the magnetic structure with spin orientation of magnetic sublattice has been shown in Fig. \textbf{2.} In the present work, we have investigated magnetic and magnetocaloric properties of GdCrTiO$_5$ compound.  GdCrTiO$_5$ has chosen as a low-temperature refrigerant material for two main reasons: (i) the large angular momentum of localized $4f$ shell electrons of Gd$^{3+}$ ($J$$=$7/2) and (ii) very low antiferromagnetic transition temperature ($T_N$) of Gd sublattice. Magnetic and thermodynamic properties suggest that GdCrTiO$_5$ is a frustrated magnet.\cite{bas} Due to strong frustration, the Gd moments order at very low temperature, $T_N$$=$0.9 K, whereas the Cr moments do not show any long-range ordering.\cite{tb,eb} To the best of our knowledge, there is no report on large MCE for such type of exchange frustrated magnets. Also, GdCrTiO$_5$ is electrically insulating and magnetization does not show thermal and field hysteresis which are important criteria in magnetic refrigerant.\\

\begin{figure}
\includegraphics[width=0.5\textwidth]{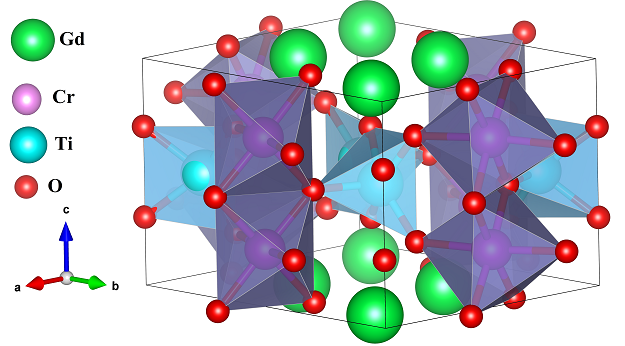}
\caption{The orthorombic crystal structure of GdCrTiO$_5$. The polyhedra formed by chromium (Cr) and titanium (Ti) with oxygen (O) atoms are shown schematically.}
\end{figure}

 \begin{figure}
\includegraphics[width=0.5\textwidth]{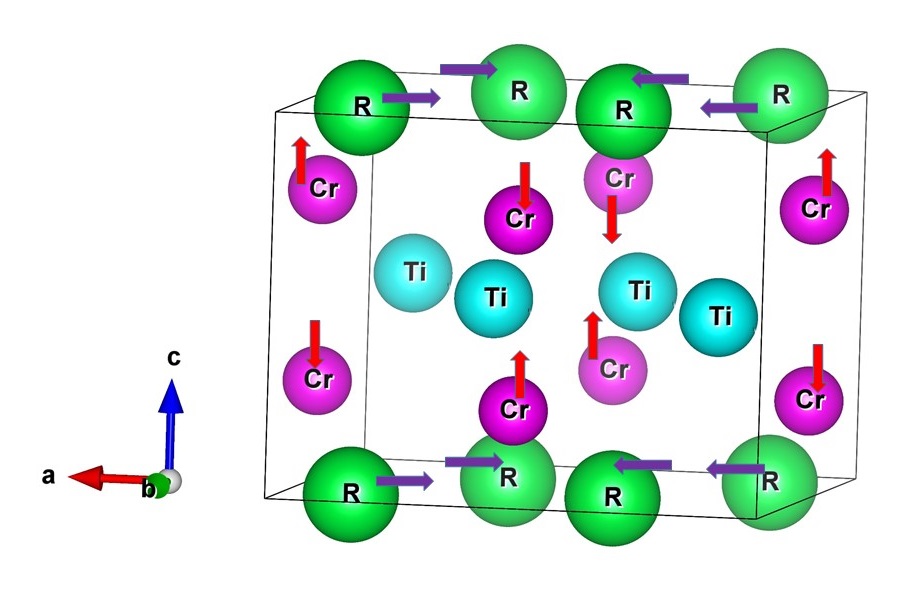}
\caption{Magnetic structure of $R$CrTiO$_5$ and the arrow indicates the orientation of magnetic moment of $R^{3+}$ and Cr$^{3+}$. In GdCrTiO$_5$, the Gd moments order antiferromagnetically below 0.9 K but  the Cr moments do not show any long-range ordering.}
\end{figure}

\begin{figure}
\includegraphics[width=0.5\textwidth]{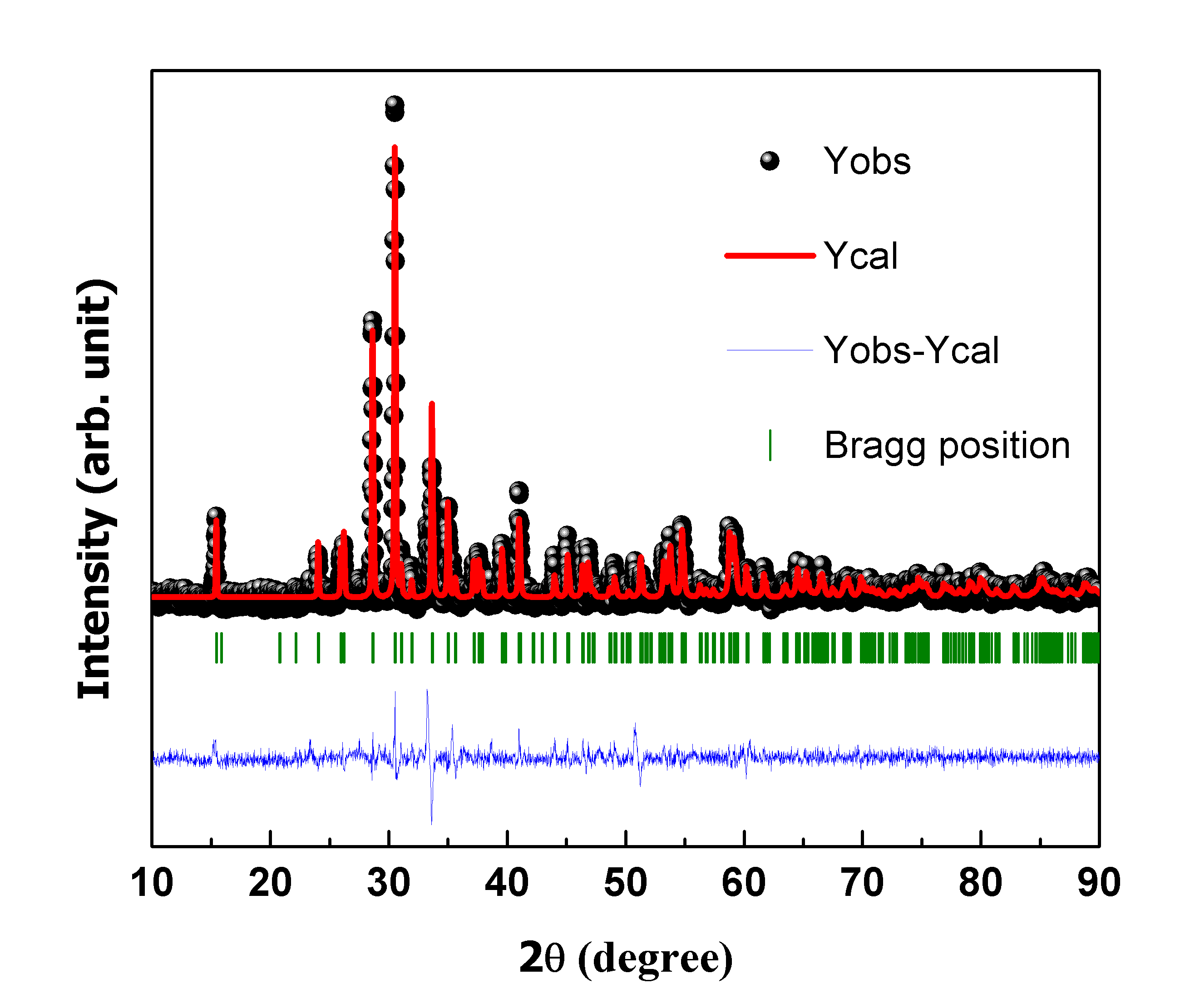}
\caption{The X-ray diffraction pattern of the polycrystalline powder of GdCrTiO$_5$. }
\end{figure}

\textbf{II. EXPERIMENTAL DETAILS}

The polycrystalline  GdCrTiO$_5$ sample was prepared by conventional solid-state reaction method using high purity Gd$_2$O$_3$ (99.9$\%$), Cr$_2$O$_3$ (99.9$\%$) and TiO$_2$ (99.9$\%$) powders. Before use, Gd$_2$O$_3$ was pre-heated at 900$^\circ$C for 24 h. Well-mixed powders of Gd$_2$O$_3$, Fe$_2$O$_3$ and TiO$_2$ in a stoichiometric ratio 1:1:2 was heated at 1250$^\circ$C for few days with intermediate grindings. Finally, the green-colored GdCrTiO$_5$ sample was reground, pressed into pellets under hydrostatic pressure and sintered at 1400 $^\circ$C for 24 h in air.

The phase purity and crystal structure of the sample have been determined by high-resolution x-ray powder diffraction with CuK$_\alpha$ radiation (Rigaku TTRAX II) ($\lambda=$1.5406 {\AA}) at room temperature. The Rietveld refinement was used for the structural analysis of diffraction pattern of powdered GdCrTiO$_5$ sample with FULLPROF software. The experimental x-ray intensity profile along with the theoretical fit and the Bragg positions are shown in Fig. \textbf{ 3.} All the peaks in the diffraction pattern can be indexed well with orthorhombic unit cell having Pbam crystallographic symmetry. Within the x-ray resolution, we did not observe any peak due to the impurity phase. The lattice parameters determined from the Rietveld profile analysis are $a=$7.4162, $b=$8.5862, and $c=$5.7807 {\AA}, which are very close to the previously reported values.\cite{eb}

A small piece of rectangular shape sample was cut from the polycrystalline pellet  for the magnetization ($M$) measurements in a SQUID vibrating sample magnetometer (Quantum Design). The data have been recorded for the isothermal magnetization measurement in the field range of 0-7 T at different temperatures between 2 and 35 K, and the temperature dependence of magnetization was measured in the range 1.8-300 K. The heat capacity measurement was performed in a physical property measurement system (Quantum Design) by relaxation method down to 1.8 K.
\\

\textbf{III. RESULTS AND DISCUSSION }

The magnetization  of GdCrTiO$_5$ has been measured as a function of temperature. The main panel of Fig. \textbf{4} shows $M$($T$) curve for an applied field 100 Oe. $M$ increases with decrease in $T$ but no clear signature of long-range magnetic ordering is observed down to the lowest measured temperature. This behavior of GdCrTiO$_5$ is quite unusual because the Cr$^{3+}$ moments in NdCrTiO$_5$ compound exhibit long-range antiferromagnetic (AFM) ordering below 20 K and due to the exchange interaction between Cr$^{3+}$ sublattice and Nd$^{3+}$ moments, the Nd$^{3+}$ moments  order at lower temperature, 13 K. \cite{sa, mg, hw, sp} On the contrary, the significantly larger moment of Gd$^{3+}$ ($S$=7/2) as compared to Cr$^{3+}$ ($S$=3/2) causes a strong spin fluctuation and tries to suppress the ordering of Cr$^{3+}$ sublattice.  For better understanding the nature of magnetic ground state of GdCrTiO$_5$, the inverse susceptibility ($\chi^{-1}$) has been plotted as a function of temperature in the inset of Fig. \textbf{4.} At high temperature above 150 K, susceptibility ($\chi$) follows the Curie-Weiss law, $\chi$=$C/(T-\theta$$_{CW}$), where $C$ is the Curie constant and $\theta$$_{CW}$ is the Curie-Weiss temperature. From the linear fit to the high temperature data, we have calculated the values of effective paramagnetic moment $\mu$$_{eff}$$=$8.8 $\mu$$_B$/f.u.  and $\theta$$_{CW}$$=$-24 K.  These values match well with the previously reported ones \cite{bas}. Negative $\theta$$_{CW}$ suggests that the dominating exchange interaction in GdCrTiO$_5$ is AFM in nature. The above value of $\mu$$_{eff}$ is very close to the theoretical one (8.83 $\mu$$_B$/f.u), determined using the two-sublattice model, $\mu$$_{eff}$=[$(\mu_{eff}^{Gd})^2+(\mu_{eff}^{Cr})^2]^{1/2}$,  where $\mu_{eff}^{Gd}$ (=7.94 $\mu$$_B$/Gd) and $\mu_{eff}^{Cr}$(=3.87 $\mu$$_B$/Cr) are respectively, the values of effective moment of  Gd$^{3+}$ and Cr$^{3+}$ ions in the PM state.\\

\begin{figure}
\includegraphics[width=0.5\textwidth]{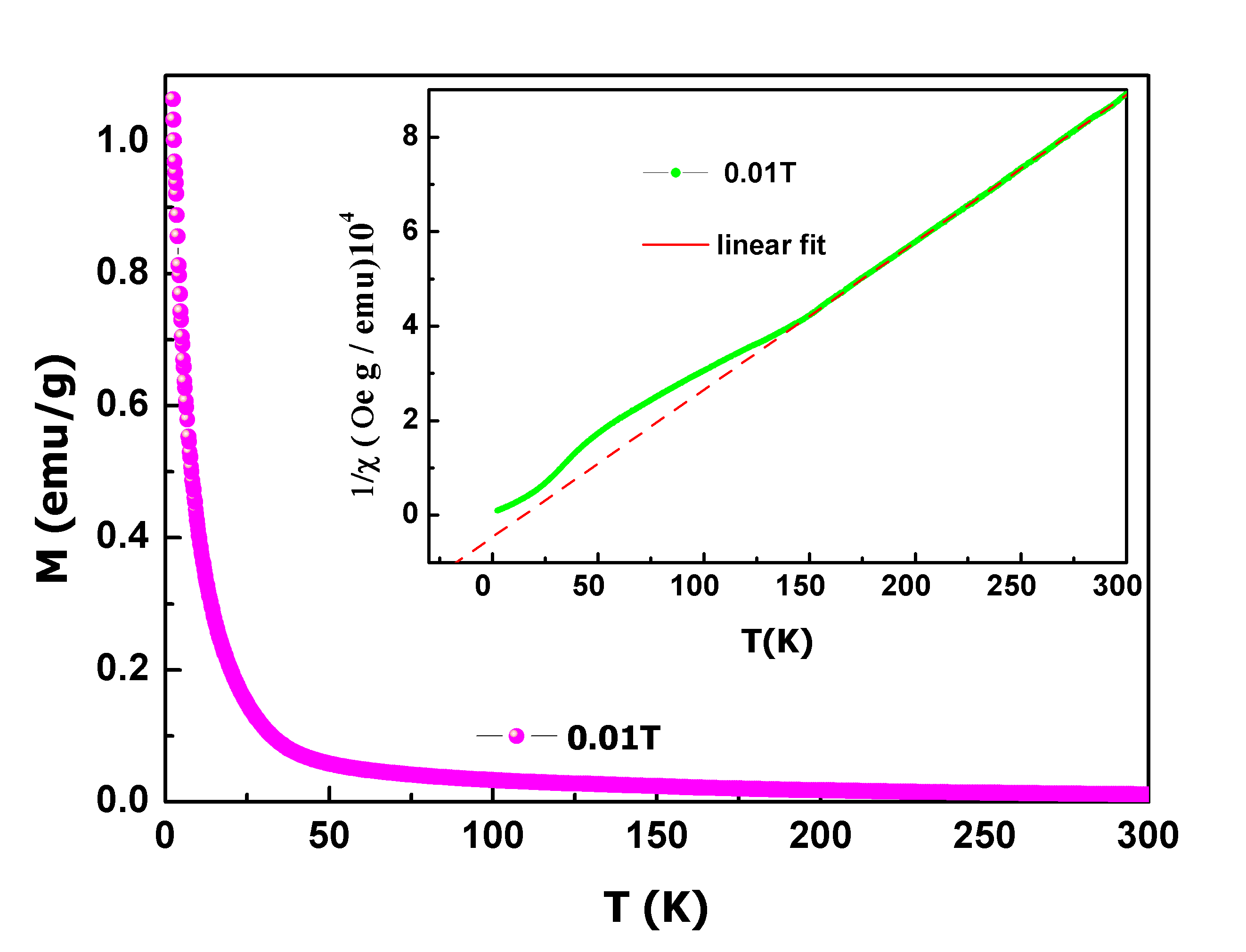}
\caption{The main panel shows the temperature dependence of magnetization for GdCrTiO$_5$ at 0.01T and the inset shows the Curie-Weiss fit at high temperature.}
\end{figure}

\begin{figure}
\includegraphics[width=0.5\textwidth]{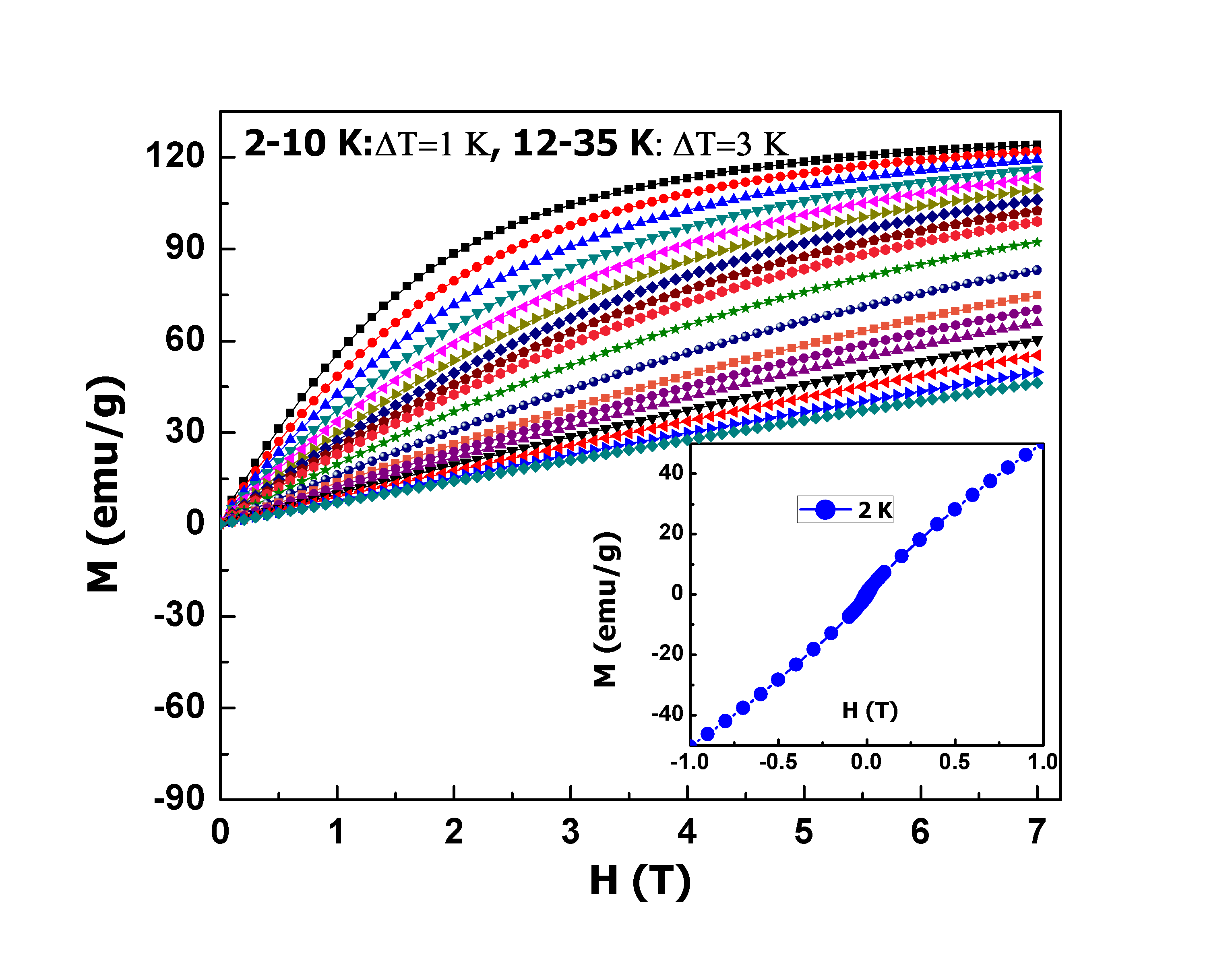}
\caption{The main panel shows the isothermal magnetization plots for GdCrTiO$_5$ in the temperature range of 2-35 K and the inset shows the hysteresis loop at 2 K in the low-field region.}
\end{figure}

With decreasing temperature, $\chi^{-1}$ starts to deviate from the linear behavior below $\sim$150 K, which is quite high. Such a nonlinear behavior of $\chi^{-1}$($T$) curve at high temperatures well above the $\theta$$_{CW}$ may be due to the strong spin fluctuations in the PM state of the system. Several geometrically frustrated rare-earth transition metal oxides  exhibit strong spin fluctuations due to their nearly triangular network of the magnetic ions.\cite{sato,katsufuji,katsufuji1,barn} As a result, the long-range AFM ordering in these compounds occurs at a much lower temperature than the deduced Curie-Weiss temperature; the magnetic energy scale of the system. The reduction in AFM transition temperature, $T_N$, is a signature of frustration and the value of the ratio $\theta_{CW}/T_N$ can be used as a measure of the spin frustration strength.\cite{sato,katsufuji,katsufuji1} The spin system is classified as the one with strong geometrically frustrated, if this ratio exceeds 10, because the simple mean-field theory fails to explain such a huge reduction in $T_N$ \cite{katsufuji1}. In hexagonal manganites ($R$MnO$_3$), the maximum value of $\theta_{CW}/T_N$ is reported to be $\sim$10 and these systems are considered to be strongly frustrated ones \cite{sato,katsufuji,katsufuji1}. As the present system orders antiferromagnetically below 0.9 K and the observed values of $\theta$$_{CW}$ are within 24-33 K, \cite{bas,eb} we find 26$<$$\theta_{CW}/T_N$$<$37, which is significantly larger than the value reported for the hexagonal manganites. Thus, GdCrTiO$_5$ can be considered as a strongly spin frustrated system like several other multiferroics.  \\

\begin{figure}
\includegraphics[width=0.5\textwidth]{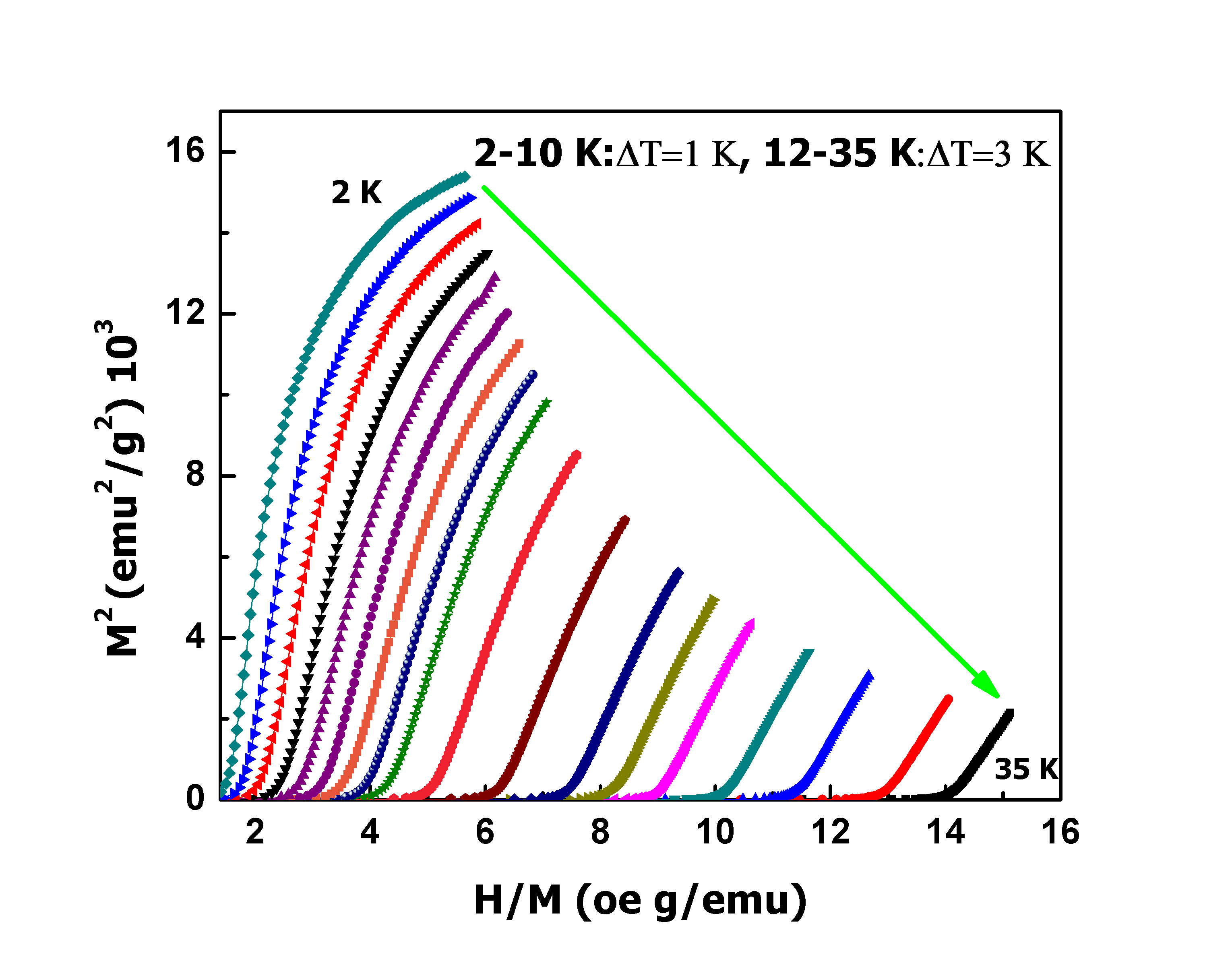}
\caption{The Arrott plots for the GdCrTiO$_5$.}
\end{figure}

In order to explore the influence of applied magnetic field on magnetic ground state, we have measured the field dependence of magnetization  in GdCrTiO$_5$  up to 7 T at different temperatures in the range 2-35 K. The field dependence of $M$ is shown in the main panel of Fig. \textbf{5.} As in the case of a typical ferromagnet, $M$ increases monotonically with the increase in $H$ and tends to saturate at high field and low temperature. At 2 K and 7 T, the  value of magnetic moment is about 7.4 $\mu_B$/f.u which is about 6$\%$ higher than the spin only moment of Gd, indicating a small contribution from the Cr sublattice. The inset of Fig. \textbf{5} shows the low-field $M$($H$) curve of the present compound. $M$($H$) does not display any hysteresis. The evolution of $M$ with $H$ indicates that the field-induced transition is second-order in nature. The nature of magnetic phase transition is important in refrigeration technology. The second-order magnetic phase transition is always preferable than the first-order. Generally, second-order phase transition  exhibits very low or no hysteresis, whereas the first-order transition may exhibit significant hysteresis loss which is undesirable in magnetic cooling technology. In order to understand the exact nature of the field-induced magnetic phase transition, the $M$($H$) curves have been transformed into well-known Arrott plots. Fig. \textbf{6} shows $M^{2}$ versus $H/M$ plots for GdCrTiO$_5$. The slope of $M^{2}$ versus $H/M$ curve is useful to determine the order of both temperature and field driven magnetic phase transition. The positive slope of the $M^{2}$ versus $H/M$ suggests that the field-induced phase transition in GdCrTiO$_5$ compound is second-order continuous in nature. \\

\begin{figure}
\includegraphics[width=0.5\textwidth]{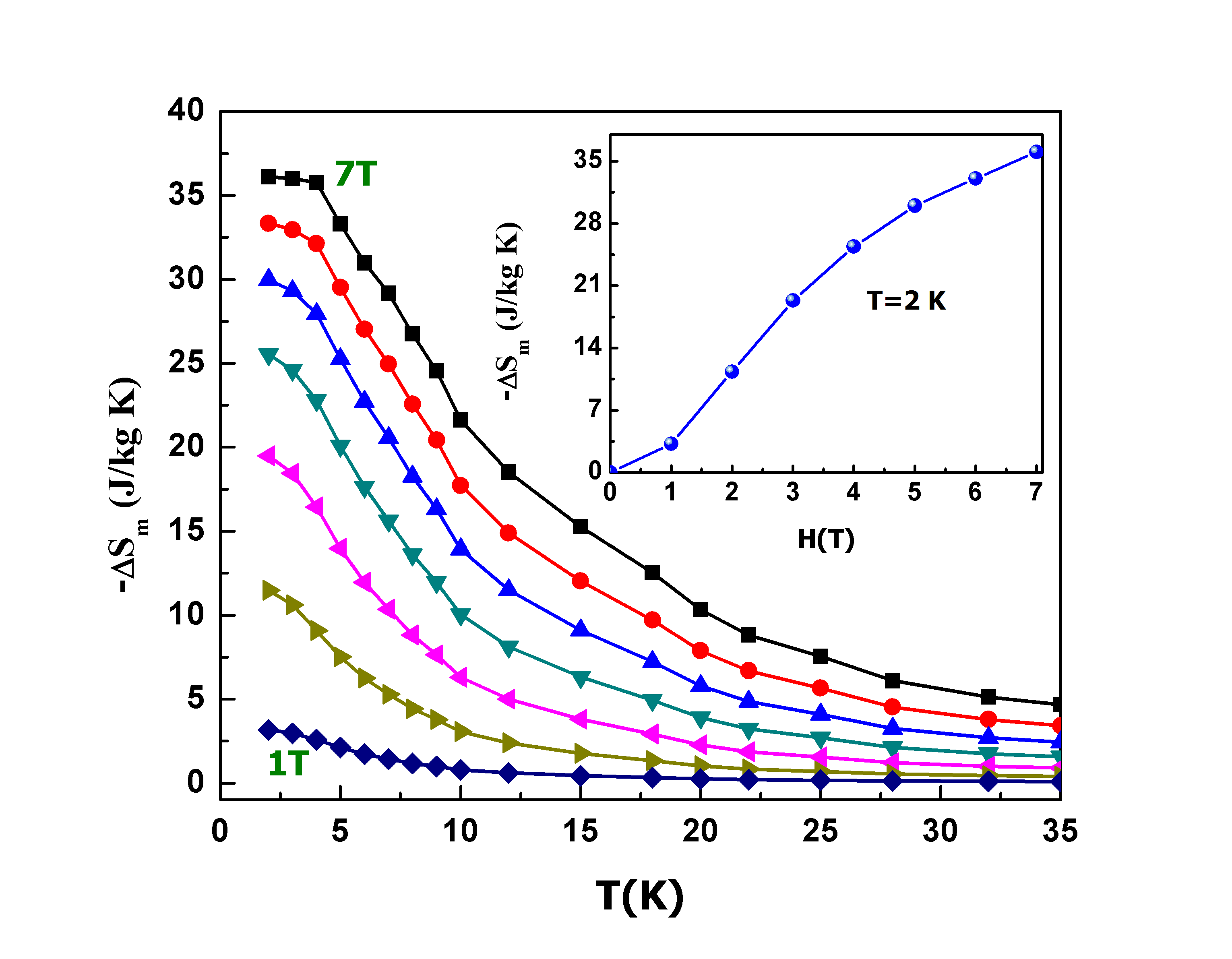}
\caption{ The temperature variation of $\Delta S_{\rm m}$ for GdCrTiO$_5$ calculated from the magnetization data and inset shows the field dependence of $\Delta S_{\rm m}^{max}$ at 2 K.}
\end{figure}

The large field-induced isothermal magnetization indicates giant magnetic entropy change in GdCrTiO$_5$. In order to test whether this material is suitable for magnetic refrigeration in the low-temperature region, the magnetic entropy change has been calculated using the Maxwell equation, $\Delta S_m=\int_{0}^{H}(dM/dT)dH$. As the magnetization measurements are done at discrete field and temperature intervals, $\Delta S_{\rm m}$ is numerically calculated using the following expression,
\begin{eqnarray}
\Delta S_{\rm m} = \sum_{i}\frac{M_{i+1} - M_{i}}{T_{i+1} -T_i} \Delta H_i
\end{eqnarray}
where $M_i$ and $M_{i+1}$ are the magnetic moments at temperatures $T_i$ and $T_{i+1}$, respectively for a change in magnetic field $\Delta H_{\rm i}$. The temperature dependence of $\Delta S_{\rm m}$ has been calculated from the magnetic field dependence of magnetization data at different temperatures using the above equation. The temperature dependence of $\Delta S_{\rm m}$ for field variation up to 7 T has been shown in Fig. \textbf{7.} $\Delta S_{\rm m}$ is found to be very large and negative down to the lowest measured temperature. The maximum value of $\Delta S_{\rm m}$ (-$\Delta S_{\rm m}^{max}$) increases with increase in field and reaches as high as 36 J kg$^{-1}$ K$^{-1}$ for a field change of 0-7 T, which is more than double of the previously reported values of $\Delta S_{\rm m}$ for other members of the $R$Mn$_2$O$_5$ family.\cite{bali, jan, heng} We would also like to mention that the observed value of $\Delta S_{\rm m}$ is significantly larger than that reported for several rare-earth transition metal oxides and intermetallic compounds.\cite{rm,li,ros,ba} A comparative study of $\Delta S_{\rm m}$ of GdCrTiO$_5$ with other rare-earth based oxide compounds at same field and temperature range has been shown in Table \textbf{I.} \\

\begin{figure}
\includegraphics[width=0.5\textwidth]{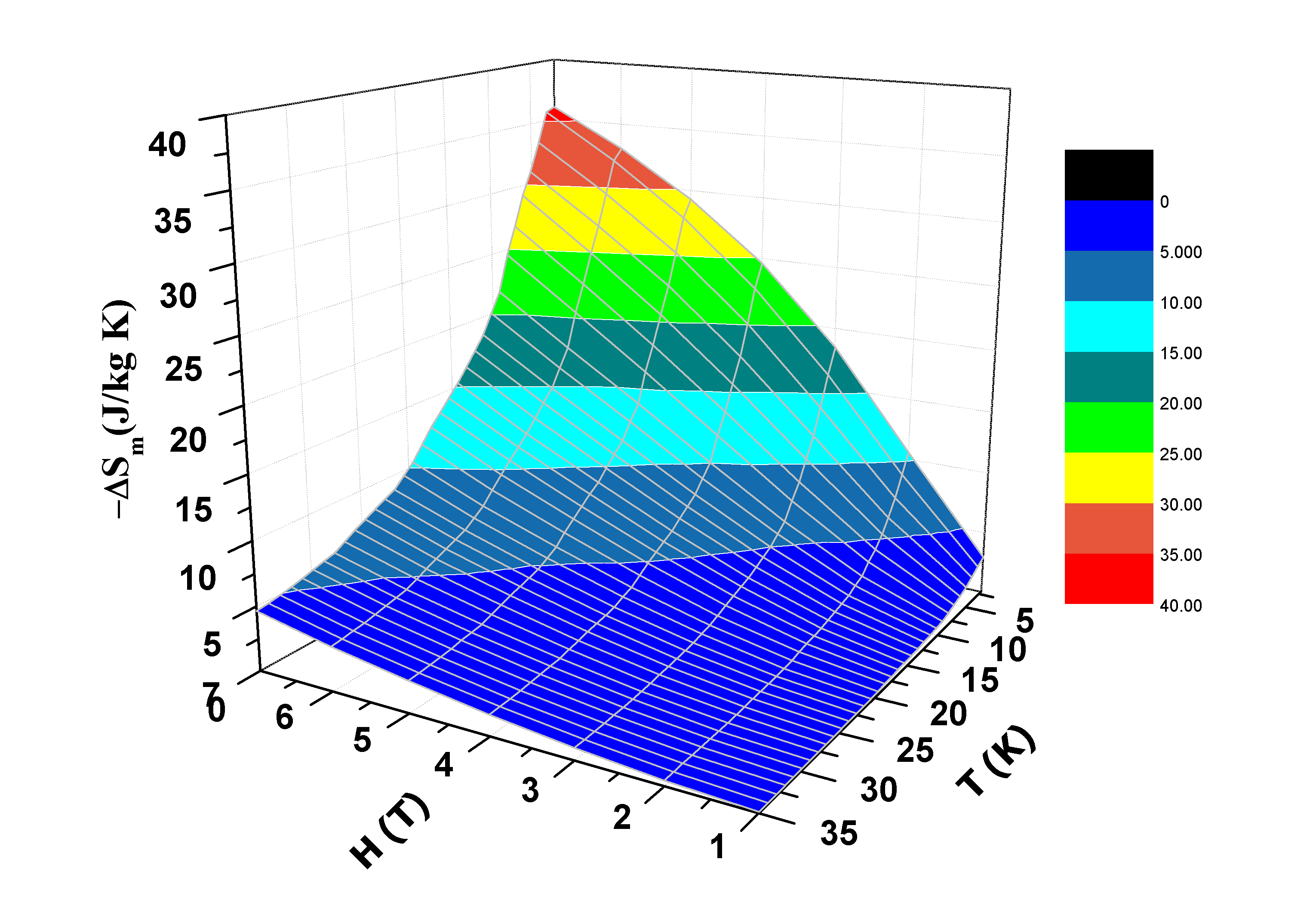}
\caption{ The contour plot of $\Delta S_{\rm m}$ as functions of temperature and magnetic field for GdCrTiO$_5$.}
\end{figure}

Apart from the value, the nature of temperature dependence of $\Delta S_{\rm m}$ is very important for magnetic refrigeration.  In a typical ferromagnet or antiferromagnet,  $\Delta S_{\rm m}$ increases with decreasing $T$ in the PM state but it decreases rapidly below the onset of long-range ordering temperature, i.e., $\Delta S_{\rm m}$ decreases on the both sides of $T_C$ or $T_N$. In this context, it may be noted that undoped and doped EuTiO$_3$, EuDy$_2$O$_4$ and GdVO$_4$ exhibit huge MCE at low temperature.\cite{pm,md,sr,mi,kd} The values of $\Delta S_{\rm m}^{max}$ in these compounds are also comparable to the present system. However,  for the above mention compounds, $\Delta S_{\rm m}$($T$) shows a strong decrease in the low-temperature region. On the other hand,  $\Delta S_{\rm m}$ in the present system does not  decrease down to 2 K but a saturation-like behavior appears below 5 K for fields above 5 T. For application, $\Delta S_{\rm m}$ should be reasonably large at low or moderate magnetic field strength. The field dependence of $\Delta S_{\rm m}$ is displayed in the inset of Fig. \textbf{7} at 2 K.  From the contour of $\Delta S_{\rm m}$ with temperature and magnetic field as shown in Fig. \textbf{8} one can see that  $\Delta S_{\rm m}$ is quite large even at low field. For example, the values of  $\Delta S_{\rm m}$ at 2 K are 12 and 20 J kg$^{-1}$ K$^{-1}$ for field change of 2 and 3 T, respectively which can be achieved using a permanent magnet. Another remarkable feature of low-field $\Delta S_{\rm m}$($T$) curve is that  instead of saturation behavior at low temperature, $\Delta S_{\rm m}$ increases with decrease in $T$ (d$\Delta S_{\rm m}$/d$T$$<$0). So, $\Delta S_{\rm m}$ can be significantly large even in the subkelvin region.\\

We believe that the large MCE and the unusual temperature dependence of $\Delta S_{\rm m}$ are associated with infinite degenerate magnetic frustrated ground states. The frustration in the present system occurs due to the competition between the Dzyaloshinskii-Moriya interaction and the spatial anisotropy exchange interaction which is one of the main characteristics of $R$Mn$_2$O$_5$ series. With application of magnetic field, the degeneracy in the ground state tends to lift and causes the frustrated magnetic moments to polarize in the field direction, as a result, large magnetic entropy change has been occurred. Theoretical investigation shows that the enhancement of magnetocaloric effect is related to the presence of a macroscopic number of soft modes in frustrated magnets below the saturation field.\cite{me} However, there are very few experimental reports to support such theoretical prediction to obtain large MCE in frustrated systems.\cite{ss,sp}  \\

For application perspective, the mechanical efficiency ($\eta$) is an important parameter for magnetic refrigerant cycle. So, we have also calculated $\eta$ for the present GdCrTiO$_5$ system and compared with different magnetocaloric materials as shown in Table \textbf{I} using the method described by Moya et al.\cite{moya} From the Table \textbf{ I}, one can see that the efficiency of GdCrTiO$_5$ is larger as compared to several magnetic refrigerants in the same temperature region and magnetic fields.\\

\begin{table}
\caption{Comparison of magnetic entropy change($\Delta S_{\rm m}$) and mechanical efficiency ($\eta$) of different magnetocaloric materials with respect to GdCrTiO$_5$}
\vspace{1.5em}
\centering
\begin{tabular}{|l|c|c|c|c|c|c|}
 \hline
Materials&$T_{0}(K)$&$\Delta H_{0}$(T)&$\Delta S_{\rm m}$(J kg$^{-1}$ K$^{-1}$)&$\eta$(\%)&Ref.\\
\hline
GdCrTiO$_5$ &5&2(5)&7.7(25.1)&51(37)&this work\\
\hline
ErFeO$_3$ &5&2(5)&3(12)&47(19)&\cite{hua} \\
\hline
TbCrO$_3$ &5&2(5)&4(12)&20(17)&\cite{lhy}\\
\hline
HoMnO$_3$ &5&2(5)&1(3)&19(18)&\cite{midya} \\
\hline
DyMnO$_3$ &5&2(5)&0.2(3.8)&3(2)&\cite{midya} \\
\hline
HoMn$_{2}$O$_5$ &5&2(5)&1(4)&15(7.4)&\cite{bali}\\
\hline
GdVO$_4$ &5&2(5)&0.27(1.06)&44(40)&\cite{kd} \\
\hline
EuDy$_{2}$O$_4$ &5&2(5)&8(20)&30(20)&\cite{mi}\\
\hline
\end{tabular}
\vspace{1em}

T$_0$, operating temperature; $\Delta H_{0}$, change in applied magnetic field;Data for $\Delta H_{0}$ = 5 T are presented parenthetically.
\end{table}

\begin{figure}
\includegraphics[width=0.5\textwidth]{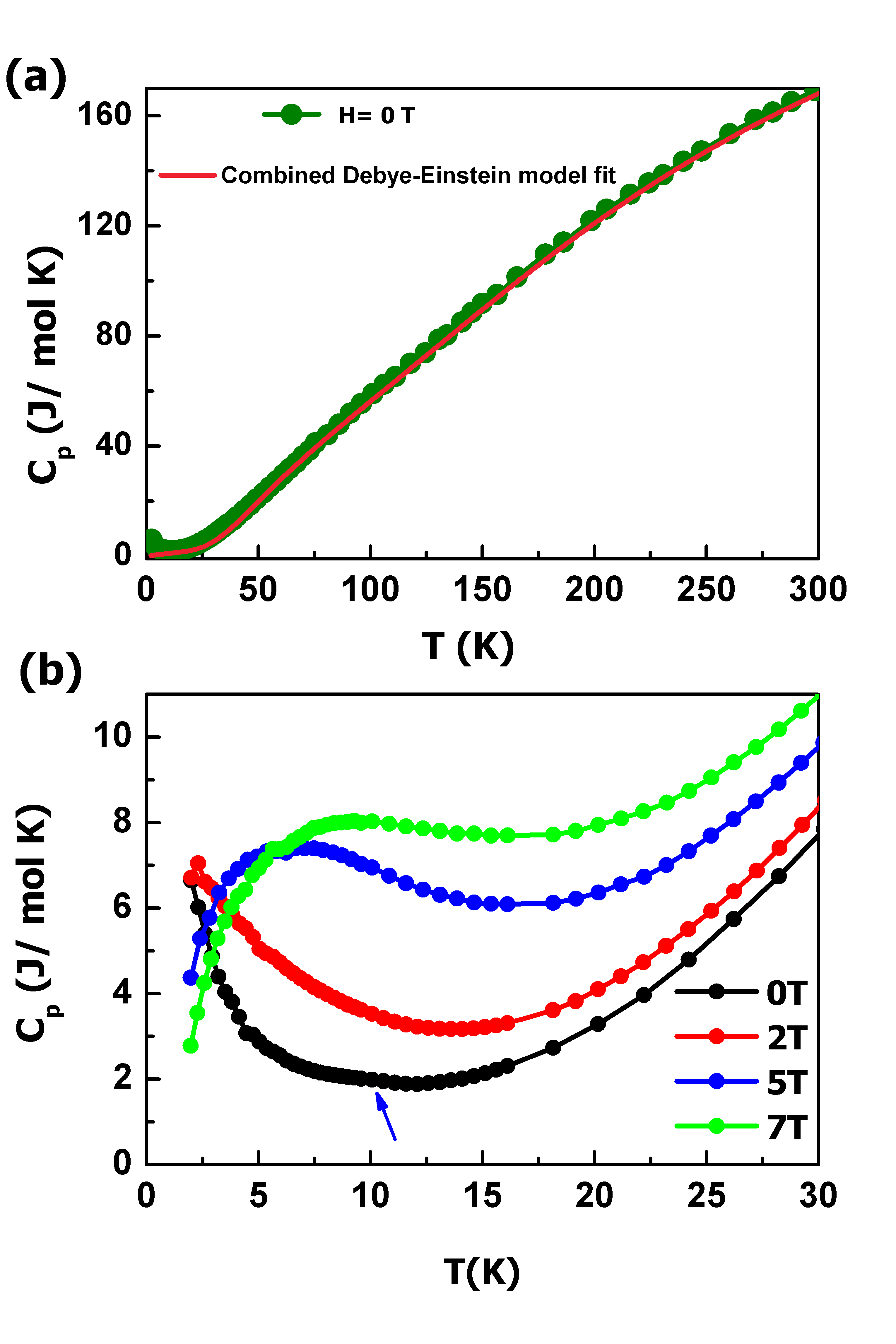}
\caption{ (a) The zero-field heat capacity data for GdCrTiO$_5$ compound and the solid line is the combined Debye-Einstein fit. (b) The field dependence of heat capacity for GdCrTiO$_5$. The arrow indicates the short-range ordering of chromium near 10 K.}
\end{figure}

For further insight into the magnetic ground state, we have measured the heat capacity ($C_p$) of GdCrTiO$_5$. Fig. \textbf{9(a) and 9(b)} show the temperature dependence of specific heat at different applied fields. At zero field as shown in Fig. \textbf{9(a)}, initially $C_p$ decreases with decreasing $T$ down to 12 K and then increases with further decrease in $T$. No strong anomaly due to the long-range magnetic ordering has been observed within the measured temperature range 1.8-300 K.  However, a careful observation reveals an extremely weak anomaly at temperature around 10 K, which is just above our experimental resolution. Similar weak anomaly in $C_p$ has been reported earlier and attributed to short-range AFM ordering of Cr$^{3+}$ spins.\cite{eb}  The nature of anomaly suggests that the transition is very weak and the entropy associated with this transition is negligible. The increase of $C_p$ at low temperature indicates the onset of long-range ordering of Gd$^{3+}$ sublattice below 1.8 K. Indeed, temperature dependence of specific heat reveals a pronounced peak at 0.9 K due to the Gd moment ordering. \cite{eb}  With the application of magnetic field, the nature of low-temperature $C_p$($T$) curve changes drastically. Up to 2 T, $C_p$ enhances with increase in field strength without showing any peak but a broad peak appears around 7 K at 5 T which shifts towards higher temperature with further increase of field.  The zero-field $C_p$($T$) curve can be fitted well with the combined Debye plus Einstein model over a wide temperature range as shown in Fig. \textbf{9(a)}. At low temperature, however, the fitted curve  deviates from the observed experimental data. The obtained lattice heat capacity calculated using the Debye plus Einstein model fitting, was subtracted from the total heat capacity to determine the magnetic contribution ($C_m$). The magnetic entropy $S_m$ is obtained by integrating ($C_m/T$)d$T$. However, due to the absence of magnetic ordering within the measured temperature range and the sharp increase of $C_p$ at low temperature, magnetic entropy cannot be determined for zero and 2 T fields. For this reason, $S_m$ has been calculated from the high-field $C_p$($T$) curves such as at 5 and 7 T where the peak appears well above 2 K and $C_p$ is small at low temperature. At high temperature, the entropy is expected to be close to the  full saturated value Rln(2J+1)$=$17.2 J kg$^{-1}$ K$^{-1}$ for the Gd$^{3+}$. Fig. \textbf{10} shows that $S_m$ starts to saturate above 15 K and the saturated value is close to  17.2 J kg$^{-1}$ K$^{-1}$ for both $H$$=$5 and 7 T. At high temperatures well above $T_N$, the saturated value of entropy should nearly be the same for all fields even for the zero magnetic field. Comparing the deduced value of zero-field $S_m$ with that for 7 T field, we find that at zero field, a significant amount of entropy (10.6 J kg$^{-1}$ K$^{-1}$) is released just below 1.8 K. So, this extra amount was  added to the  zero-field entropy data to determine $S_m$ for 0 T. For 2 T, the corresponding value is 7.6 J kg$^{-1}$ K$^{-1}$.  As the maximum normalized entropy ($S_m$)/R is very close to 2, we conclude that a major fraction of 4f spins of Gd$^{3+}$ is taking part in the magnetic ordering. We have also calculated the zero-field magnetic entropy from the reported heat capacity data at low-temperature (0.05-20 K) and observe that the obtained value is very close to that of ours.\cite{eb}\\

To check the consistency in our results on magnetic entropy change estimated from $M$($H$) data, $\Delta S_{\rm m}$ has also been calculated independently from the field dependence of heat capacity using the relation $\Delta S_{\rm m}$=$\int_{0}^{T}[C_p(H_2,T)−-C_p(H_1,T)]/TdT$, where $C_p$($H,T$) is the specific heat at a field $H$. $\Delta S_{\rm m}$ as calculated from the heat capacity data is shown in the inset of Fig. \textbf{10} for different magnetic fields as a function of temperature. It is clear from the plots that the values of $\Delta S_{\rm m}$  estimated from the heat capacity data are close to that calculated from  magnetization.  For an example,  the calculated value of $\Delta S_{\rm m}^{max}$ from magnetization is 30.2 J kg$^{-1}$ K$^{-1}$ whereas that from the heat capacity data is 27 J kg$^{-1}$ K$^{-1}$ for the same field change 0-5 T. The small difference in the value of $\Delta S_{\rm m}^{max}$ may be due to underestimation of the magnetic heat capacity. \\

\begin{figure}
\includegraphics[width=0.5\textwidth]{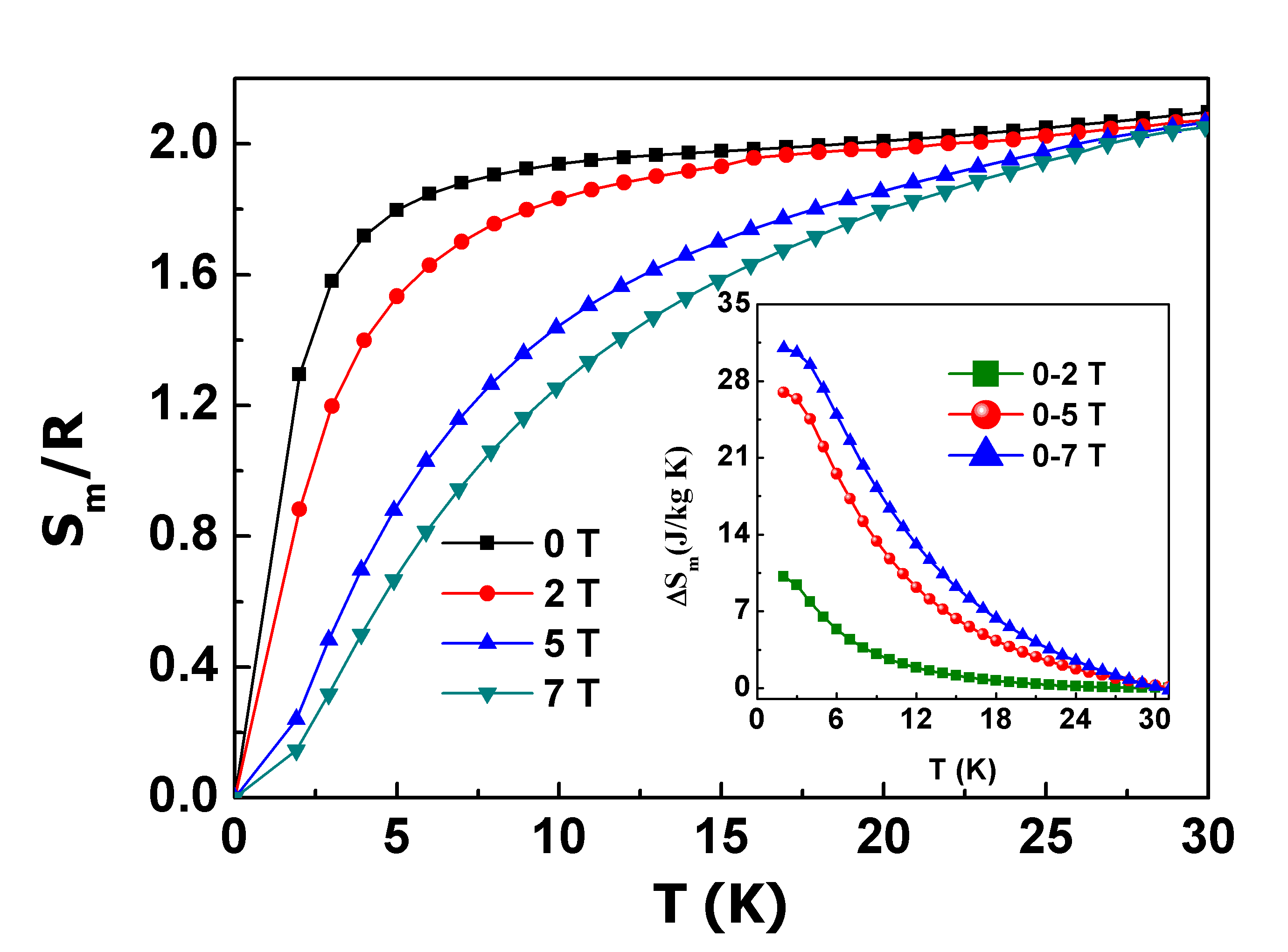}
\caption{ The temperature variation of magnetic entropy with field for GdCrTiO$_5$  compound and the inset shows the variation of $\Delta S_{\rm m}$ calculated from the heat capacity data. }
\end{figure}

\begin{figure}
\includegraphics[width=0.5\textwidth]{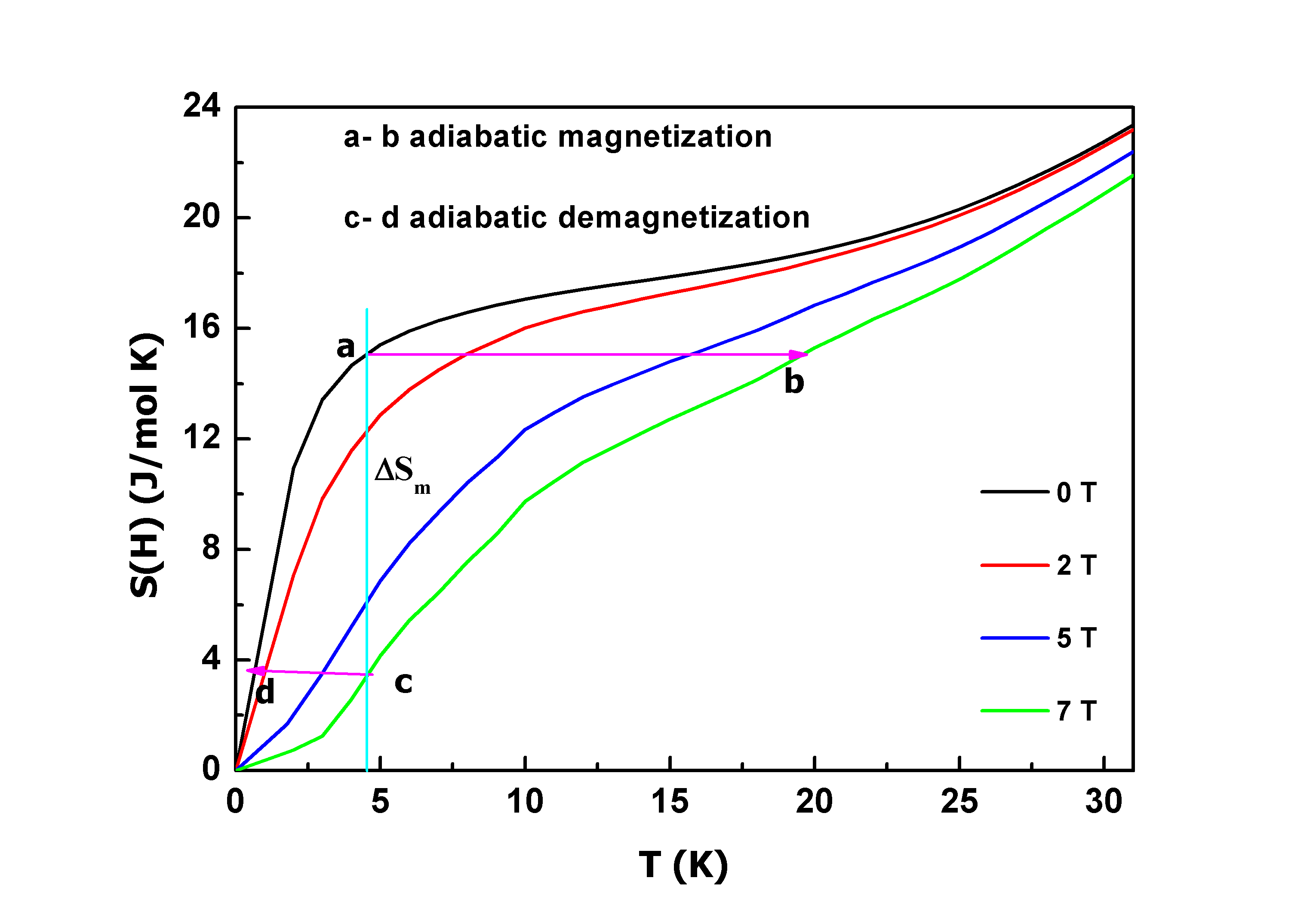}
\caption{ The variation of entropy at different fields. The horizontal arrow from $a$ to $b$ indicates the adiabatic heating and $c$ to $d$ indicates the adiabatic cooling. Whereas the vertical arrow indicates the isothermal entropy change for the magnetic field change 0-7 T.}
\end{figure}

\begin{figure}
\includegraphics[width=0.5\textwidth]{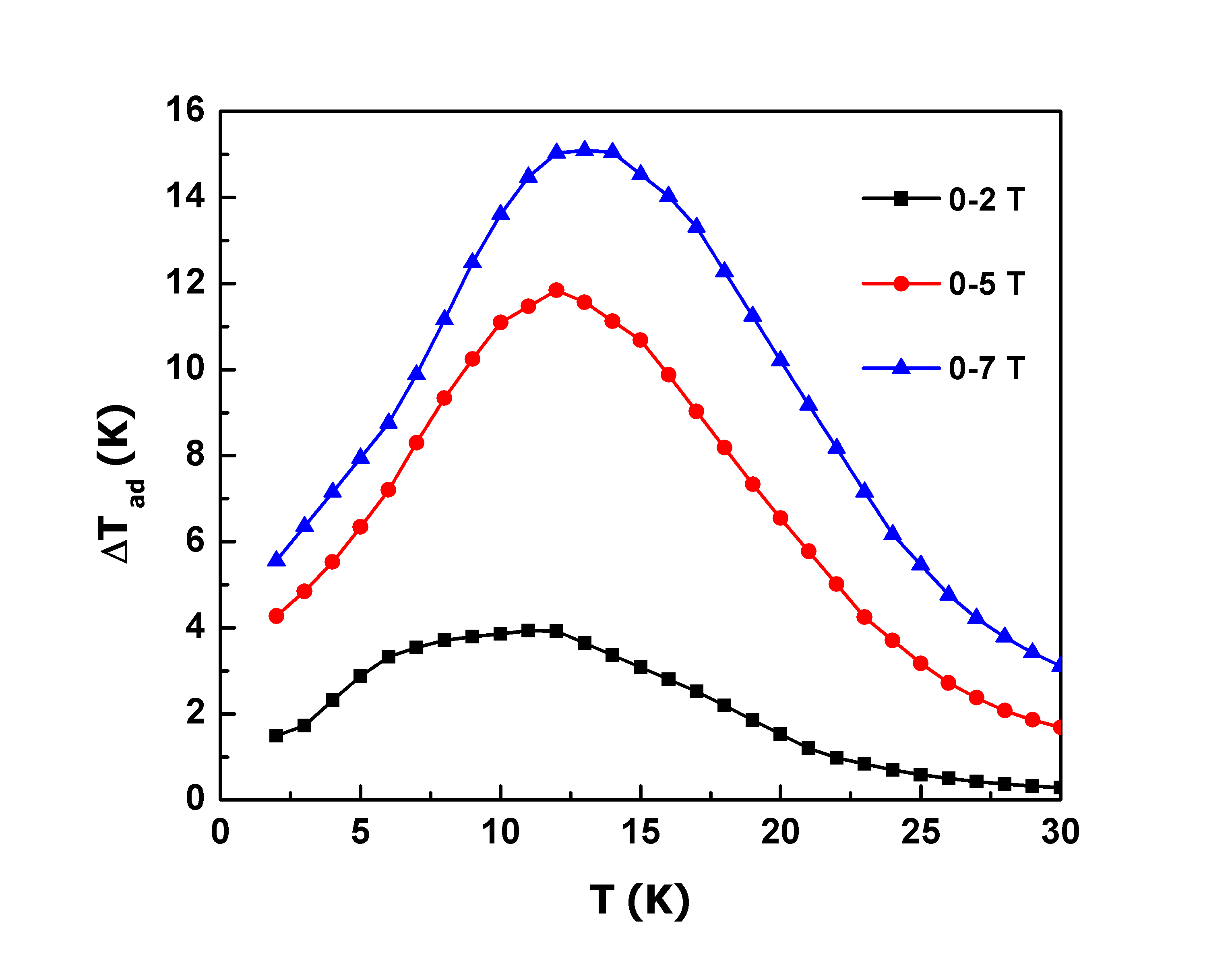}
\caption{The temperature dependence of $\Delta T_{\rm ad}$ for GdCrTiO$_5$ at different magnetic fields.}
\end{figure}

Another very important parameter related to the magnetic refrigeration is $\Delta T_{\rm ad}$  which is the isentropic temperature difference between $S$($H, T$) and $S$(0, $T$).  For this, we have calculated the entropy $S$($H, T$) at field $H$ after subtracting $\Delta S_{\rm m}$($H, T$) determined using the heat capacity data, from the zero-field entropy $S$(0,$T$). The variation of entropy at different fields  has been shown in Fig. \textbf{11.} The temperature dependence of $\Delta T_{\rm ad}$ is shown in Fig. \textbf{12.} The maximum value of $\Delta T_{\rm ad}$ reaches as high as 15 K at 7 T. Thus, both $\Delta S_{\rm m}$ and $\Delta T_{\rm ad}$ are large in GdCrTiO$_5$ system. Similar to  $\Delta S_{\rm m}$, $\Delta T_{\rm ad}$ is also quite large at low and moderate field strength.  However, there is an asymmetry in the $\Delta T_{\rm ad}$($T$) curve near 10 K, when applying a field adiabatically ($\Delta T_{\rm ad}$ heating) and removing the field adiabatically ($\Delta T_{\rm ad}$ cooling). The entropy increases rapidly in zero applied field but it increases at a slower rate in presence magnetic field. So to interpret the deduced values of adiabatic temperature change, in Fig. \textbf{12}, we have shown the actual heating ($a$ to $b$ arrow) and cooling ($c$ to $d$ arrow) effects due to adiabatic magnetization and adiabatic demagnetization, respectively. These two processes explain the difference between cooling and heating cycles in the magnetocaloric effect of GdCrTiO$_5$.\\

\textbf{IV. SUMMARY}

In summary, we have studied the magnetic and magnetocaloric properties of GdCrTiO$_5$  through magnetization and heat capacity measurements. In GdCrTiO$_5$, magnetocaloric parameters are quite large. The maximum values of isothermal entropy change and adiabatic temperature change are 36 J kg$^{-1}$ K$^{-1}$ and 15 K, respectively at 7 T. This compound also demonstrates a remarkable magnetocaloric effect even at low and intermediate applied fields. Unlike several potential low-temperature magnetic refrigerants, $\Delta S_{\rm m}$ in the present compound does not decrease at low temperature. Our result suggests that GdCrTiO$_5$ could be a potential material for magnetic refrigeration at low temperature.\\

\textbf{ACKNOWLEDGEMENT}

The authors would like to thank A. Pal for technical help during sample preparation and measurements.\\

\end{document}